\documentclass[prb,aps,twocolumn,superscriptaddress,showpacs]{revtex4}
\usepackage[dvips]{graphicx}
\usepackage{amsmath}
\usepackage{epsfig}
\def\k{{\bf k}}
\begin{document}

\title{  Probing the pairing symmetry of the iron pnictides with electronic Raman scattering}

\author{G. R. Boyd}
\affiliation{Department of Physics, University of Florida,
    Gainesville, FL 32611, USA}
\author{T. P. Devereaux}
\affiliation{
Stanford Institute for Materials and Energy Sciences, 
SLAC National Accelerator Laboratory, 2575 Sand Hill Road, Menlo Park, CA 94025}
\affiliation{
Geballe Laboratory for Advanced Materials, Departments of Physics and Applied Physics, Stanford University, CA 94305.}
\author{P. J. Hirschfeld}
\affiliation{Department of Physics, University of Florida,
    Gainesville, FL 32611, USA}
\author{V. Mishra}
\affiliation{Department of Physics, University of Florida,
    Gainesville, FL 32611, USA}
\author{D. J. Scalapino}
\affiliation{Department of Physics, University of California,
Santa Barbara, CA 93106-9530 USA   }

\begin{abstract}
An important issue in the study of the iron-arsenic based superconductors is the symmetry of the
 superconducting gap, a problem complicated by multiple gaps on different Fermi surface
 sheets. 
Electronic Raman scattering is a flexible bulk probe which allows
one in principle to determine gap magnitudes and test for gap
nodes in different regions of the Brillouin zone by employing
different photon polarization states.  Here we calculate the clean
Raman intensity for
 $A_{1g}$,  $B_{1g}$ and $B_{2g}$ polarizations, and discuss the peak structures and low-energy power
 laws which might be expected for several popular models of the
 superconducting gap in these systems.  

\end{abstract}

\maketitle

\section{Introduction}

Since their discovery\cite{ref:kamihara}, there has been a
considerable effort to understand the origin and nature of
superconductivity in iron-pnictide materials (for early reviews
see Refs. \onlinecite{Sadovskii:2008,aoki:2008,ref:Mazin}).  Initial
information on the structure of the gap is often provided by power
laws in the temperature dependence of thermodynamic and transport
properties, which are related to the topology of the
superconducting gap in its nodal regions. Nuclear magnetic
resonance  (NMR) studies
\cite{ref:RKlingeler,ref:Grafe,ref:Ahilan,ref:Nakai}  showed a
low temperature $T^3$ spin lattice relaxation rate typical of a  gap with
nodes.  
However, penetration depth measurements
\cite{ref:Hashimoto,ref:Malone,ref:Martin,ref:Hashimoto2,ref:Gordon,ref:Gordon2,ref:Fletcher}
have been fit both to exponential activated $T$-dependence,
indicative of a fully gapped state, and low-$T$ power laws. 
ARPES measurements on single crystals of 122-type
materials
\cite{ref:Zhao,ref:Ding,ref:Kondo,ref:Evtushinsky,ref:Nakayama,ref:Hasan}
measured the spectral gap reporting isotropic or nearly isotropic gaps on
all Fermi surface sheets. It is
possible that these differences reflect genuinely  different
ground states in different materials. However, the complex
interplay of multiband effects, unconventional
pairing, and disorder leaves open the possibility that a 
single ground state symmetry exists in all
the Fe-pnictides, and that differences in measured properties can
be understood by accounting for electronic structure differences and
disorder\cite{ref:Vorontsov,ref:vivek}. It is likely that a
consensus will be reached only after careful measurements using
various probes on the same material, and systematic disorder
studies.

We argue here that measurement of the electronic Raman scattering
in the superconducting state can provide important information on
the structure of the bulk superconducting order parameter through its sensitivity to
both symmetry and gap scales. Here we discuss the Raman scattering for some
simple models of the Fe-pnictide superconductors.
In general the energy of the peaks in the Raman intensity are
directly related to the magnitude of the gaps on the various Fermi
sheets.  However, whether a given Raman polarization weights a
given gap strongly or weakly depends on the polarization state of
the measurement via the Raman vertex $\gamma_{\bf k}$.  This is
particularly important for superconductors where the gap is
strongly momentum dependent, and was exploited successfully in
cuprate materials  to help determine the $d$-wave symmetry in
those systems\cite{DevereauxRMP,Devkampf,DevEinz,DevVirZaw}.   In addition, the presence of
nodes and the dimensionality of the nodal manifold may be
determined by comparison with low energy power laws in the Raman
intensity in different polarization states.   

\begin{figure}
\includegraphics[width=0.8\columnwidth]{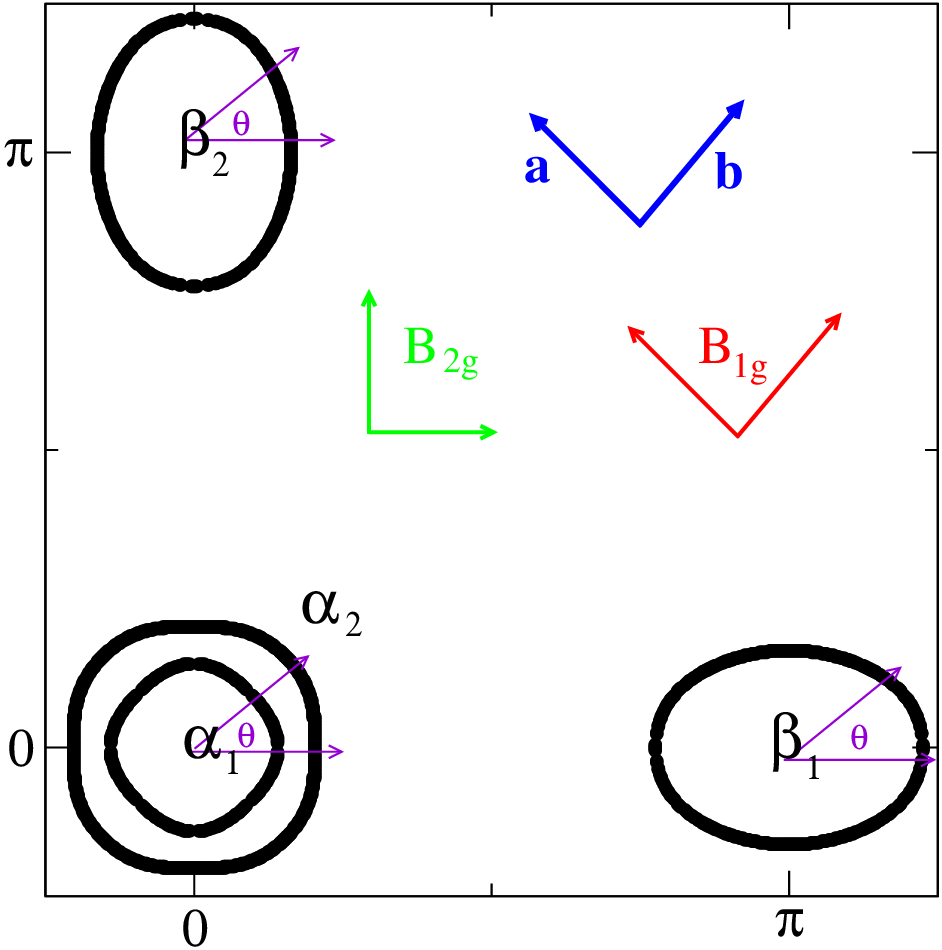}
\caption{Fermi surface of model Fe-pnictide represented in
unfolded (1-Fe) Brillouin zone. The crystalline $a, b$ axes are indicated in blue,
and the polarization geometries for incoming and scattering photon
polarizations are denoted for $B_{1g}$ and $B_{2g}$ geometries. Note that our
symmetry notation is rotated by 45 degrees with respect to the lattice symmetries.} \label{FS}
\end{figure}

In the Fe-pnictide materials, density functional theory
calculations\cite{ref:Singh,ref:Cao} for the paramagnetic state
have generally found a multisheeted Fermi surface consisting of
concentric, nearly circular hole pockets around the $\Gamma$ point
(here referred to as ``$\alpha$ sheets") and nearly circular
electron pockets around the M points, or $\tilde X$ point in the
unfolded, 1-Fe zone (referred to as ``$\beta$ sheets").  These
sheets are plotted in Fig. \ref{FS}. Spin fluctuation models of
pairing\cite{ref:Kuroki,ref:Ikeda,ref:Yanagi,ref:Graser} have
usually found that the leading pairing instability had $s$-wave
symmetry,
and noted that the next leading channel had
$d_{x^2-y^2}$  symmetry. Wang {\it et al.}~\cite{ref:Wang}, who
studied the pairing problem using a functional renormalization
group approach within a 5-orbital framework, also found that the
leading pairing instability occured in the $s$-wave channel, and
that the next leading channel had $d_{x^2-y^2}$ symmetry.
These calculations differ according to whether or not actual nodes
are found in the leading pairing states, but they all find large
anisotropies over the Fermi surface sheets, not anticipated in the
original predictions of extended-$s$-type states with isotropic
gaps which changed sign between the $\alpha$ and $\beta$
sheets\cite{ref:Mazin_exts,ref:Chubukov_exts}.

While ARPES
experiments\cite{ref:Zhao,ref:Ding,ref:Kondo,ref:Evtushinsky,ref:Nakayama,ref:Hasan}
have reported weakly anisotropic gaps, this may be due to momentum
resolution issues and the difficulty of preparing good surfaces in
these systems at present.  In particular, the samples used in
these experiments may be sufficiently dirty at the surface that
considerable momentum averaging, with concomitant gap averaging,
could be taking place\cite{ref:vivek}.  Angle-dependent  specific
heat measurements in a magnetic field\cite{ref:spht}, which probe
the bulk, may be feasible  in the future but are difficult at
present due to the requirement of large clean crystals.  Raman
scattering with use of different polarizations may therefore be
the best current method of acquiring momentum-dependent
information on the structure of the bulk superconducting order
parameter. We discuss several cases below which should allow
extraction of the crude momentum dependence and possible nodal
structure of the order parameter over the Fermi surface.  Some of
these cases have been considered in an earlier paper by Chubukov
et al.\cite{ref:Chubukov_raman}, who however focussed solely on
the $A_{1g}$ polarization, and examined vertex corrections due to
short-range interactions. However, they neglected ``charge
backflow" effect of the long-range part of the Coulomb interaction
which is required to ensure charge conservation.   Here we show
that for the $A_{1g}$ case, significant changes are to be expected
due to the coupling of the $A_{1g}$ Raman charge fluctuations due
to the backflow effects.

While the expressions for the $B_{1g}$ and $B_{2g}$ channels are
generally well-characterized by the bare bubble calculation for
the cuprates, it is well-known that the $A_{1g}$ contribution is
significantly more complicated due to the issue of charge backflow
as noted above, and to the number of different
excitons which may be pulled down from a gap edge condensate. In
systems with several condensate pairing instabilities lying in
close proximity in parameter space, strong excitonic peaks may occur. 
While the possibility of such
excitonic modes is of interest for the pnictides, in this paper we
focus on the generic quasiparticle features for Raman scattering,
and will consider excitonic peaks in a future publication. Charge
backflow and Coulomb screening is explicitly included in all
$A_{1g}$ calculations, but pair interaction 
corrections will be neglected. We will however also consider the possibility
that due to the form for the Raman vertices, which are allowed when there are
multiple Fermi surfaces, there can also be backflow effects on symmetry
channels other than $A_{1g}$ in the Fe-pnictides.

We begin in Section \ref{clean} by considering model one-band
clean systems with gaps inspired by
 proposals for the Fe-pnictides to illustrate what
intuition we can gain regarding the Raman response for various
polarizations. In Section \ref{realistic}, we consider gaps on all
four fermi surfaces predicted by density functional theory.   We
present our conclusions in Section \ref{conclusions}.

\section{ Electronic Raman Scattering in clean system}
\label{clean}

\subsection{General Theory}
\label{theory}

Raman scattering is the inelastic scattering of polarized light
from a material. For a review see Ref. \onlinecite{DevereauxRMP}.
The cross section of the scattered light is proportional to
the imaginary part of the channel-dependent Raman susceptibility
\begin{equation}
\chi_{\gamma,\gamma}(\omega)=\int_0^{\beta}d\tau e^{-i\omega_m\tau}
\langle T_{\tau}\tilde\rho_{\gamma}(\tau)\tilde\rho_{\gamma}(0)\rangle
\mid_{i\omega_m\rightarrow w+i\delta}.
\end{equation}
Here we will take a simple frequency independent form for the
Raman vertices (non-resonant scattering) and write the effective
Raman charge fluctuations in the $\gamma$-channel as
\begin{equation}
\tilde\rho_{\gamma}=\sum_{{\bf k},\sigma}\sum_{n,m}
\gamma_{n,m}({\bf k})
c_{n,\sigma}^{\dagger}({\bf k})c_{m,\sigma}({\bf k}),
\end{equation}
where $n,m$ denote band indices.
$\gamma_{n,m}({\bf k})$
defines the momentum- and polarization-dependent Raman vertices,
which may include intra- and interband transitions. Generally, the
vertex is determined by both density and current matrix elements
between the conduction band and the excited states, and has not been calculated
for even simple metals like Al. However,
the polarization geometries of the incoming and outgoing photons
impose an overall symmetry due to the way in which
excitations are created in directions determined by the electric
field oscillations, and classifications of the anisotropy of the
Raman vertices can be employed. 

In this paper, in order to focus on general features for Raman
scattering in the pnictides, we will neglect band structure
features and treat all Fermi surface sheets as circles. This
allows for a simple symmetry classifications for the Raman
vertices as has been done in the cuprates. Expanding the
polarization-dependent vertices in Fermi surface harmonics for
cylindrical Fermi surfaces,
\begin{eqnarray}
\gamma_n(\theta)^{A_{1g}} &=& a_n + b_n\cos(4\theta)\nonumber\\
\gamma_n(\theta)^{B_{1g}}  &=& c_n\cos(2\theta)\nonumber\\
\gamma_n(\theta)^{B_{2g}}&=&  d_n\sin(2\theta),
\label{vertices}
\end{eqnarray}
with angle-independent band prefactors $a_n,b_n,c_n,d_n$ setting the
overall strength of the Raman amplitudes for band $n$.
Since an  isotropic density fluctuation vanishes for $q\rightarrow
0$, it can be shown via Eq. \ref{Eq:full} that the $A_{1g}$
contribution $a_n$ to the Raman vertex is cancelled by charge
backflow and does not contribute to the scattering cross-section.
This leaves  the $\cos(4\theta)$ as the first non-vanishing
contribution in the expansion.

As shown in Fig. 1, we note that our choice of coordinates rests on the 1 Fe
unit cell, which is rotated by 45 degrees with respect to the 2 Fe unit
cell. Thus our symmetry notation is electronic and not associated with the lattice principal
directions, and therefore our classifications are 45 degrees rotated with
respect to conventional lattice classifications. Using lattice
coordinates, what we call $B_{1g}$ would be $B_{2g}$, and vice-versa. While
this might create some confusion, it is convenient to understand the interplay
of the angular dependence of the vertices and the energy gaps in the rotated 1
Fe unit cell frame, as shown in Fig. 1. This should be kept in mind however
when one discusses, e.g., electronic excitations together with lattice
excitations. Then our symmetry labels $B_{1g}$ and $B_{2g}$ would have to be interchanged.

Other forms for the Raman vertices are allowable, with the only requirement
being that they must obey the
transformation rules according to the relevant point group symmetries 
of the crystal. We note that for multi-sheeted Fermi surfaces shown in 
Fig. 1, there are different possible forms for the vertices other than 
Eq. 9. While the Raman vertices of the $\alpha$ sheets must 
transform according to Eq. 9, for the $\beta$ sheets other forms for the 
vertices could be admissable. This includes, for the $B_{1g}$ vertex for example, 
a vertex which is momentum independent on each $\beta$ sheet but of opposite
sign. For the $B_{2g}$ vertex, a Raman vertex 
which is $p$-wave like on each $\beta$ sheet (with a change of sign) would 
also be admissable. We will explore these possibilities in Sec. \ref{realistic}.

For $n$ bands crossing the Fermi level, the intraband Raman response
in the absence of Coulomb screening and charge backflow
is given by
\begin{equation}
\chi_{\tilde\rho,\tilde\rho}(\omega)=\frac{1}{N}
\sum_{\bf k}\sum_n \gamma_n({\bf k})^2\lambda_n({\bf k},\omega),
\label{Eq:full}
\end{equation}
where
\begin{equation}
\lambda_n({\bf k},\omega)=\tanh\left(\frac{E_n({\bf k})}{2k_BT}\right)
\frac{4\mid\Delta_n({\bf k})\mid^2/E_n({\bf k})}{4E_n^2({\bf k})-
(\hbar\omega+i\delta)^2}
\label{Eq:Tsuneto}
\end{equation}
is the Tsuneto function for the $n^{th}$ band, having band dispersion
$\epsilon_n({\bf k})$, energy gap $\Delta_n({\bf k})$, and
quasiparticle energy $E_n^2({\bf k})=\epsilon_n^2({\bf k})+
\Delta_n^2({\bf k})$.
Taking the imaginary part of Eq. (\ref{Eq:Tsuneto}) we then obtain
for the Raman response at $T=0$
\begin{eqnarray}
&&\rm{Im} \chi_{\tilde\rho,\tilde\rho}(\omega)=\sum_n\rm{Im}
  \chi_{\tilde\rho,\tilde\rho}^n(\omega)=
\nonumber\\
&&\sum_n\frac{\pi N_{F,n}}{\omega} \rm{Re} \int d\theta
\gamma_n^2(\theta)
\frac{\mid\Delta_n(\theta)\mid^2}{\sqrt{\omega^2-
4\mid\Delta_n(\theta)\mid^2}}.
\end{eqnarray}




Since Raman scattering probes charge fluctuations in the
long-wavelength limit, the role of the long-range Coulomb
interaction is important. Isotropic charge fluctuations are
coupled across all unit cells, and Raman scattering at low
energies must vanish due to particle number conservation, leaving
only an inelastic light scattering peak at the plasmon energy.
Screening by the long-range Coulomb interaction can be taken into
account by including couplings of the Raman charge density
$\tilde\rho$ to the isotropic density $\rho$ fluctuations and is
given by
\begin{equation}
\chi_{\tilde\rho,\tilde\rho}^{scr}=\chi_{\tilde\rho,\tilde\rho}-
\frac{\chi_{\tilde\rho,\rho}\chi_{\rho,\tilde\rho}}{\chi_{\rho,\rho}},\label{eq6}
\end{equation}
with
\begin{equation}
\chi_{\tilde\rho,\rho}=\chi_{\rho,\tilde\rho}=\frac{1}{N}\sum_n\sum_{\bf
k} \gamma_n({\bf k})\lambda_n({\bf k}),\label{eq7}
\end{equation}
and
\begin{equation}
\chi_{\rho,\rho}=\frac{1}{N}\sum_n\sum_{\bf k}\lambda_n({\bf
k}).\label{eq8}
\end{equation}
Eqs. (\ref{Eq:full}-\ref{eq8}) constitute closed form expressions
for the intraband, non-resonant contribution to the Raman
response.

It is clear that the Raman response is in general not simply
additive with respect to the response calculated from each band separately.
Incident photons can create anisotropic charge fluctuations
according to the direction of the polarization light vector, and
those charge fluctuations relax by emitting a scattered photon and
redistributing charge density via intrinsic electron scattering
mechanisms such as electron-impurity, electron-phonon,
electron-electron interactions, or via breaking of Cooper pairs.
The anisotropy of the charge fluctuations created with light can
be controlled by aligning incident and scattered photon
polarization vectors, transforming according to the elements of
the irreducible point group of the crystal. For a material with $D^{4h}$
tetragonal symmetry, and a single Fermi sheet, 
the $B_{1g}$ and $B_{2g}$ Raman responses are not coupled to the long-range
Coulomb interaction. As a consequence, the
Raman charge densities for these channels do not couple to the
pure charge density channel, and the terms given in Eq. (\ref{eq7}) vanish.
However, $A_{1g}$ fluctuations need not vanish over the unit cell,
and therefore they can couple to isotropic charge density, giving
the finite backflow represented by the second term in Eq.
(\ref{eq6}). We will see that this can also occur for the $B_{1g}$ case for
Raman vertices which are allowed when one has multiple Fermi surfaces as
formed in the Fe-pnictides.

From the expression for the Raman response we see that in the case of an isotropic
gap, $\Delta_\k=\Delta$, there should always be a peak at
$2\Delta$.   In an unconventional superconductor, depending on the
polarization, this absorption peak is replaced by a peak or other
structure at $2\Delta_0$, twice the maximum of the gap over the
Fermi surface.  The peak will be rounded by scattering, but still
provides a measure of the magnitude of the gaps in the system and
may be compared to those determined from other experiments, e.g.
ARPES and tunnelling. 
The $B_{1g}$ and $B_{2g}$ vertices in Eq. (\ref{vertices}) have
zeros in \k-space and therefore weight the part of the Brillouin
zone away from these zeros. This is observed, e.g.  in the
$d$-wave cuprates, where the sharp $2\Delta$ peak occurs in the
$B_{1g}$ channel only, while a less pronounced feature
corresponding to a change in slope occurs at the same energy in
the $B_{2g}$ channel which weights the nodal regions most
strongly. Furthermore, the existence of nodes in a gap creates low
energy quasiparticles that cause a nonzero response for all
frequencies. This is in sharp contrast to fully gapped
superconductors whose response show a sharp gap edge with no low
energy quasiparticles. 


In what follows we consider separate cases of increasing
complexity in order to display what features for Raman scattering
in a multi-band system with different gap symmetries might  be
expected generically.

\subsection{Results for some special cases} \label{results_1band}

\subsubsection{Single Fermi sheet}

\begin{figure}
\includegraphics[width=\columnwidth]{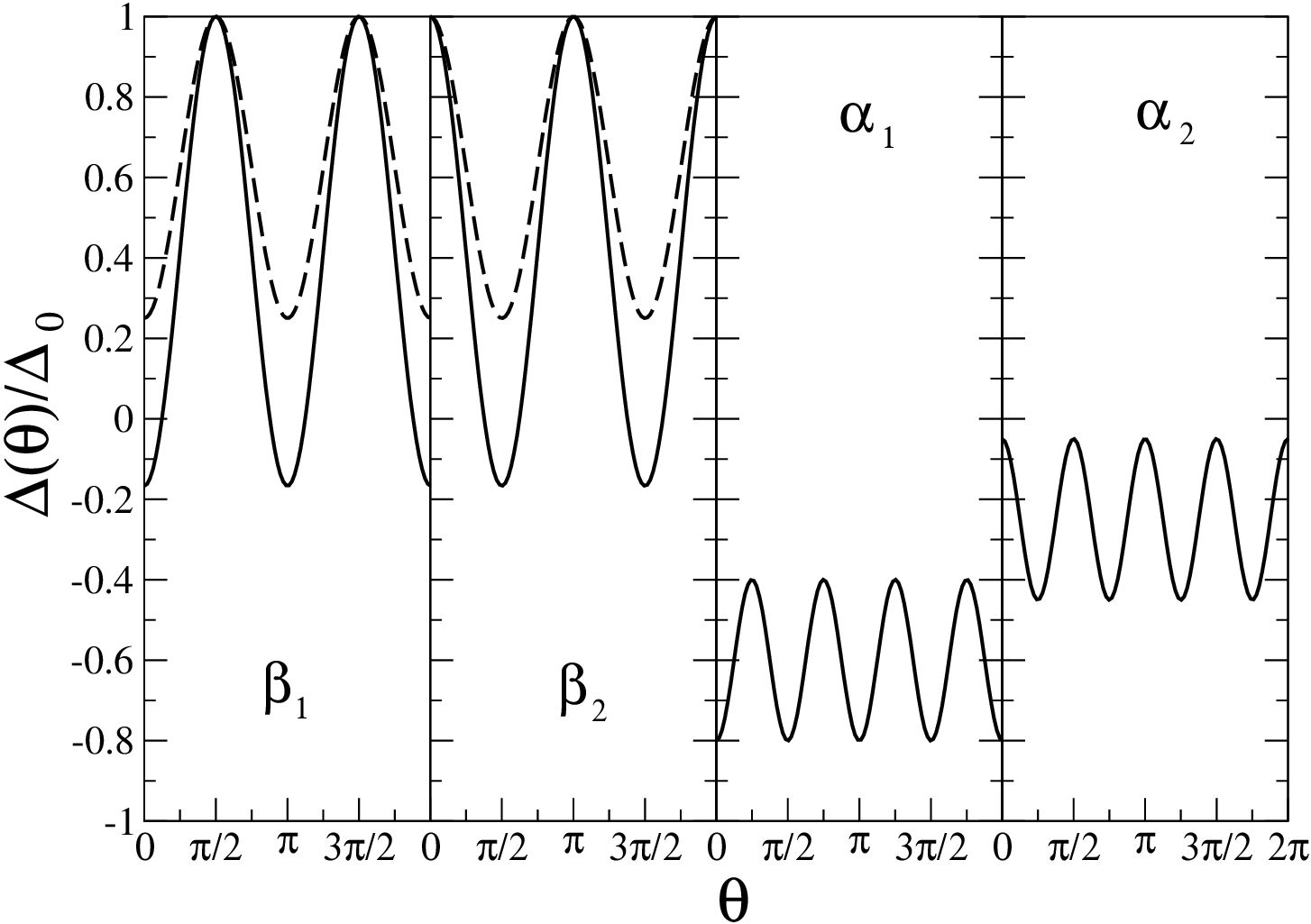}
\caption{The anisotropic energy gaps around the 4 Fermi surface
sheets as a function of angle $\theta$
  shown in Fig. 1. For the $\beta$ sheets, the solid line is for $r_\beta=1.4$ and
  the dotted line $r_\beta=0.6$ (see Eq. \ref{gaps_5band}). }
\label{Fig:2}
\end{figure}

In order to understand the type of behavior found for the multiband models
of the Fe-pnictides, it is useful to begin with some special cases. Early on,
motivated by the proximity of the Fe-pnictides to a $(\pi, 0)$
spin density wave phase and the multiple Fermi surface structure found in
LDA (local density approximation) calculations, a sign reversed $s$-wave gap
was proposed.\cite{ref:Mazin_exts}
It was suggested that spin-fluctuation scattering of electron
pairs between the $\alpha$ and $\beta$ fermi sheets could lead to pairing with
an isotropic $s$-wave gap $\Delta_0$
that changed sign between the $\alpha$ and $\beta$ Fermi surfaces. In this
case, the Raman response would consist of a peak onsetting at $2 \Delta_0$ or several peaks if
there were gaps of different magnitudes on the various Fermi
surfaces. In the absence of impurity scattering and inelastic lifetime
effects, this peak would vary as $(\omega-2\Delta_0)^{-1/2}$ as $\omega$
approaches $2\Delta_o$ from above.

Alternatively, RPA spin-fluctuation calculations and functional
renormalization group studies find anisotropic $s$-wave  gaps
which can even have nodes on the $\beta$ Fermi surfaces as well as
nearby $d$-wave  gaps with nodes on the $\alpha$ Fermi surfaces.
Here, in order to examine the Raman signatures of such states, we
consider the simple parameterization
\begin{equation}
\Delta(\theta)= \frac{\Delta_0}{1+r}  (1-r\rm{cos}(2 \theta)).
\label{Eq:gap1}
\end{equation}
This gap is plotted as a function of angle around a circular Fermi
surface.    Note, however, that for a single Fermi surface
centered at $\Gamma$, this state would not have four-fold
symmetry. Instead, the Fermi surface parameterized by $\theta$ is
intended to represent a model for the $\beta_1$ sheet of the
pnictides, and the Fermi surface angle $\theta$ is measured around
$(\pi,0)$ rather than $\Gamma$. When combined with the $\beta_2$
sheet at $(0,\pi)$, the full s-wave symmetry of the state is
restored. The gap in Eq. \ref{Eq:gap1} is normalized such that
$\Delta_0$ is the maximum over the Fermi surface, and plotted for
the values of $r=r_\beta$ shown in Fig. \ref{Fig:2}. For $r>1 $
the state has nodes on the Fermi surface, and for small values of
$r-1$ these nodes move towards 0 and $\pi$. For $r<1$, there are
no nodes but for any nonzero $r$ one has an anisotropic gap.

\begin{figure}
\includegraphics[width=\columnwidth]{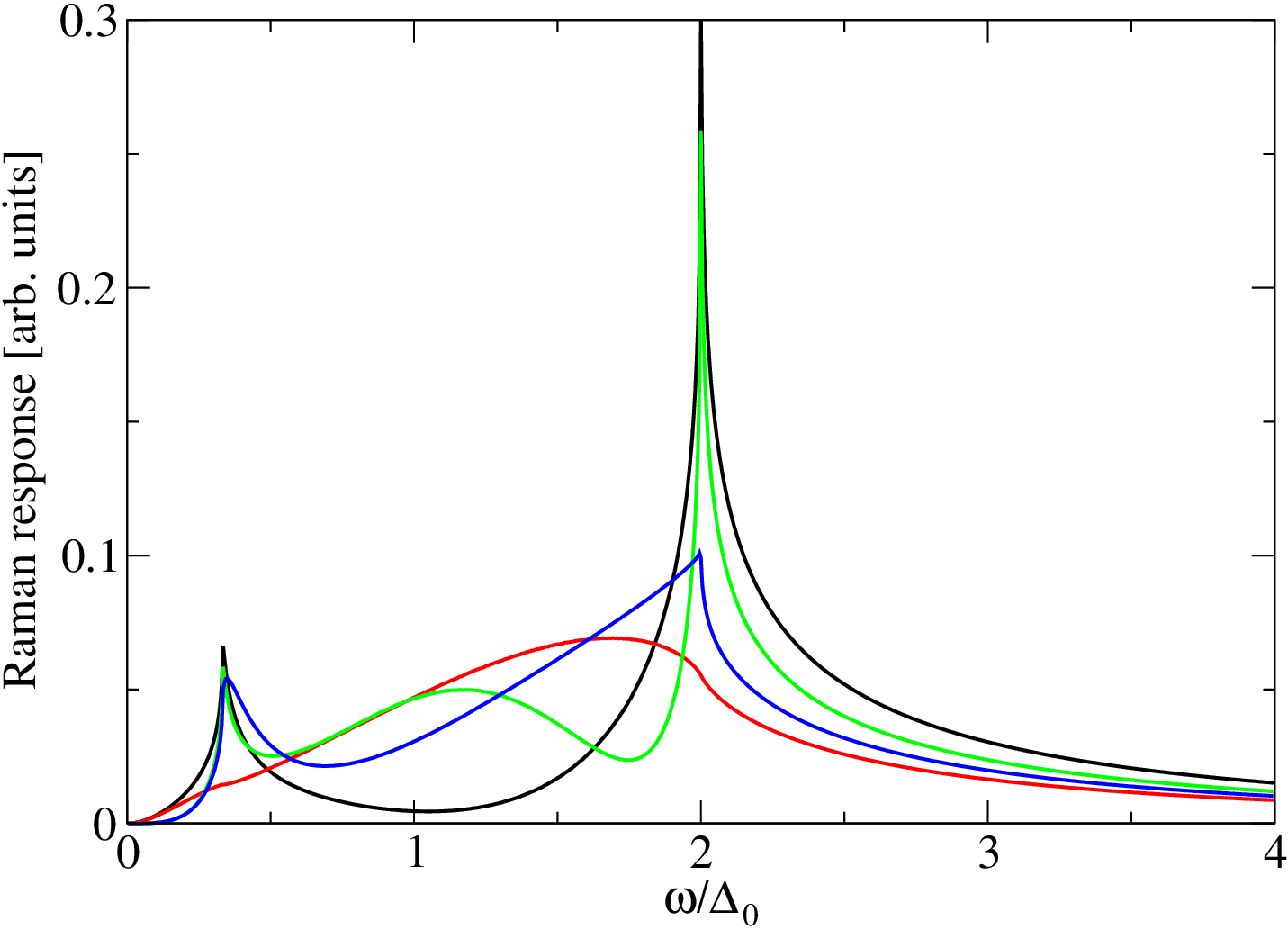}
\caption{The  Raman response of the state with energy gap Eq. \ref{Eq:gap1}
for $r= 1.4$. Black/red lines denote $B_{1g}/B_{2g}$, and green/blue
is the unscreened/screened $A_{1g}$, respectively. Note that our symmetry
classifications are according to the geometry shown in Fig. 1. For lattice
classifications, $B_{1g}$ and $B_{2g}$ should be interchanged.}
\label{Fig:3}
\end{figure}

\begin{figure}
\includegraphics[width=\columnwidth]{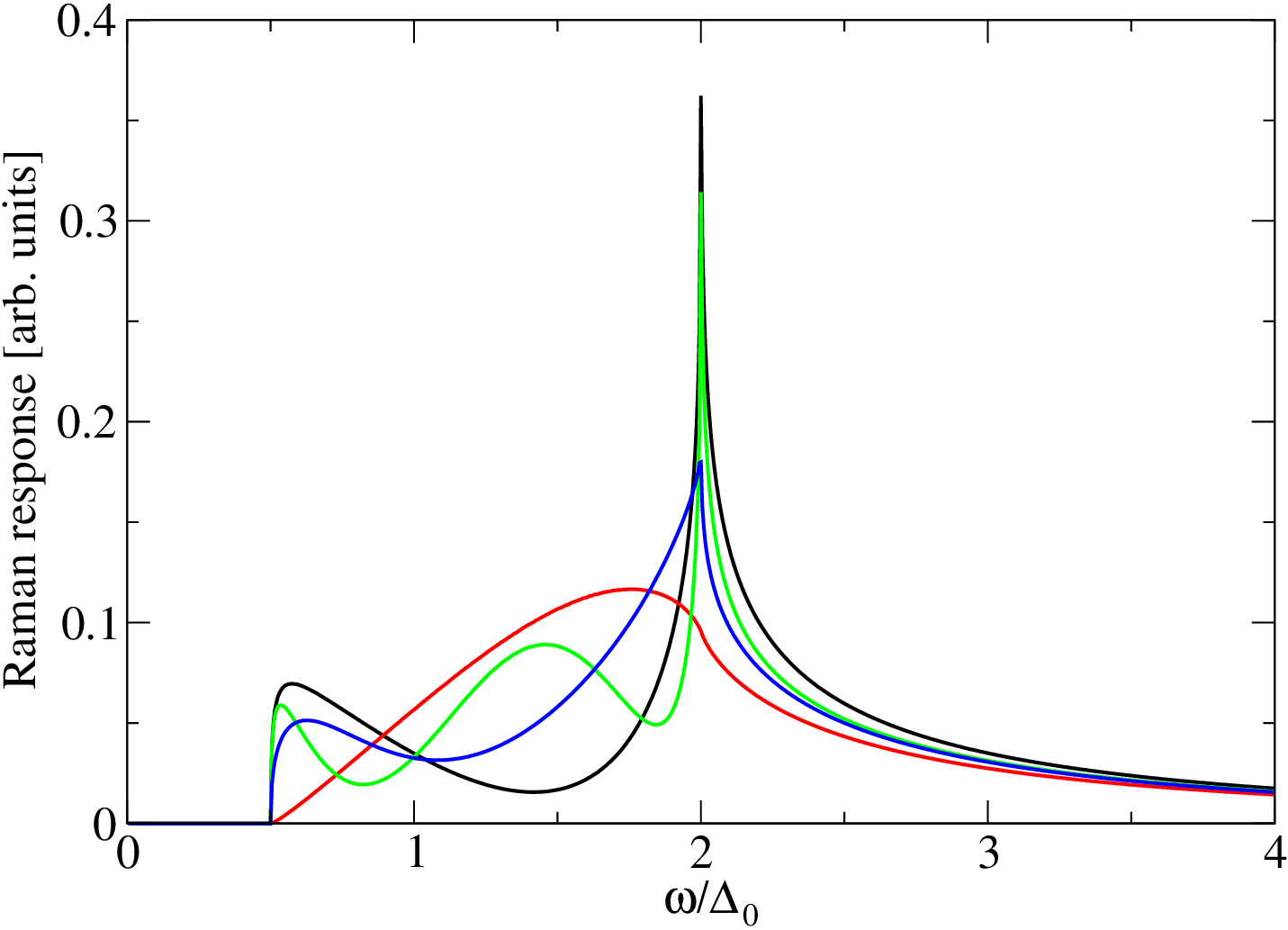}
\caption{The  Raman response of the state with energy gap Eq. \ref{Eq:gap1}
for $r= 0.6$. Black/red lines denote $B_{1g}/B_{2g}$, respectively, green/blue
is unscreened/screened $A_{1g}$.}
\label{Fig:4}
\end{figure}

In Figs. \ref{Fig:3} and \ref{Fig:4} we exhibit the $B_{1g},$
$B_{2g}$, unscreened $A_{1g}$, and screened $A_{1g}$ Raman
responses for the gaps given by Eq. (\ref{Eq:gap1}), which are
shown in the leftmost panel in Fig \ref{Fig:2}. In Fig.
\ref{Fig:3}, where $r=1.4$, there are gap nodes on the $\beta$
sheets and one finds the expected low frequency power law behavior
in which both response functions vary as $\omega$, following the
low energy behavior of the density of states, since the nodes of
the energy gap do not align with the nodes of the Raman vertices.
For the gap edge $\omega=2\Delta_{max}=2\Delta_0$, the $B_{1g}$
and unscreened $A_{1g}$ spectra have a
$\log\mid\omega-2\Delta_0\mid$ singularity. For the $B_{2g}$
response, there is a change of slope at $2\Delta_0$ since the
nodes of the vertex align with the gap maxima. A secondary $\log$
singularity appears at $\omega=2\Delta_{min}$, probed by the
$B_{1g}$ and $A_{1g}$ vertices, but not $B_{2g}$. Screening, as
pointed out before, removes all $\log$ singularities from the
unscreened $A_{1g}$ response, leaving only a cusp-like behavior at
$\omega=2\Delta_{min},2\Delta_{max}$.

For a nodeless anisotropic gap ($r=0.6$) shown as the dashed line in the
$\beta_1$ panel of Fig. \ref{Fig:2}, the Raman spectra are shown in
Fig. \ref{Fig:4}. In this case, there is a gap in the low frequency spectra.
Here, for $r<1$, $\Delta_{min}$ occurs where the magnitude of the gap has a
local minimum rather than a local maximum as it does for $r>1$. This leads to
a step discontinuity at $\omega/\Delta_0=0.5$ for the $B_{1g}$ and $A_{1g}$
spectra rather than the log singularity
seen in Fig. \ref{Fig:3} for $r=1.6$. The nodes of
the Raman vertex for the $B_{2g}$ response are aligned with $\Delta_{min}$,
leading to a linear onset at $\omega=2\Delta_{min}$ rather than a step onset.

\subsubsection{Two Fermi surface sheets}

As indicated by Eq. \ref{Eq:full}, the $A_{1g}$ Raman response for
the case of a multi-sheet Fermi surface is not in general additive
due to the Coulomb interactions and associated charge backflow.
While no such backflow appears in the $B_{1g}$ or $B_{2g}$
channels for the present case, we will see in Sec. III that for Raman vertices
which are allowed for the Fe-pnictides, there can be Coulomb backflow
contributions to the $B_{1g}$ channel. To illustrate this we calculate the $A_{1g}$ Raman
response for two Fermi surface sheets, where the energy gaps
$\Delta_{1,2}(\Theta)$ are both proportional to
$(1+r\cos(4\theta))/(1+r)$, with $r=0.2$.  For simplicity, we take the gap maximum
on one Fermi surface to be $\Delta_0$ and on the other
$\Delta_0/2$. We also take equal Raman vertices and density of
states for each band.

\begin{figure}
\includegraphics[width=\columnwidth,angle=0]{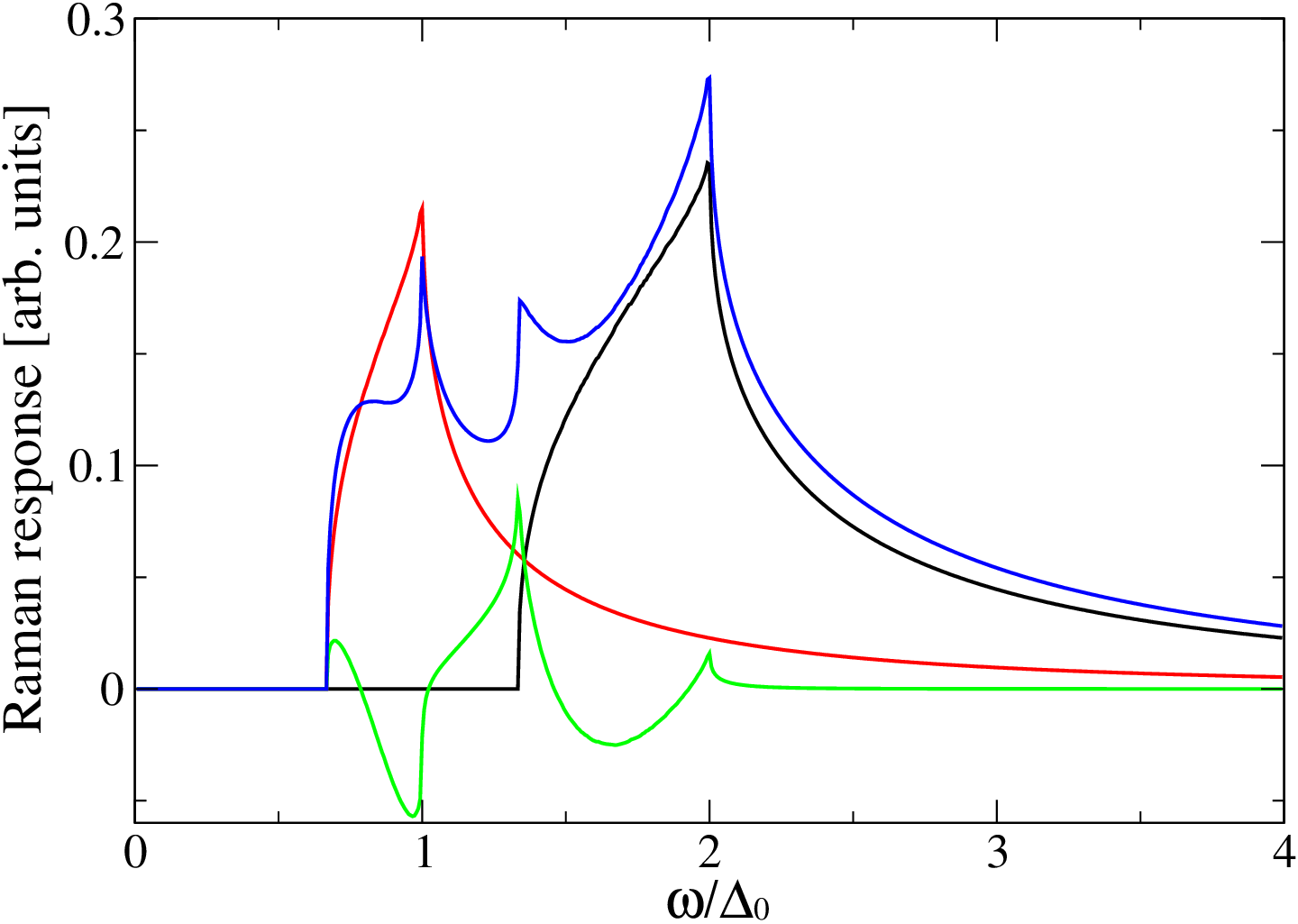}
\caption{ The screened $A_{1g}$ Raman response for 2 bands with
  $\Delta_{1,2}(\theta)$ proportional to $(1+r\cos(4\theta))/(1+r)$ with
  $r=0.2$, and the
  maximum gap on Fermi surface 1 equal to $\Delta_0$ and on Fermi surface 2
  equal to $\Delta_0/2$. ( black = screened $A_{1g}$ for band 1, red =
  screened $A_{1g}$ for band 2, green = mixing term, blue = total response.)}
\label{Fig:5}
\end{figure}

Expanding Eqs. (\ref{Eq:full}-\ref{eq8}) for the case of 2 bands,
the overall Raman response can be written as
\begin{equation}
\rm{Im} \chi_{sc}(\omega)= \rm{Im} \chi_1(\omega) + \rm{Im} \chi_2(\omega) +\rm{Im}
\Delta\chi(\omega),
\end{equation}
where
\begin{eqnarray}
\chi_{1,2}(\omega)=\frac{1}{N}\sum_{\bf k}\gamma_{1,2}^2({\bf
  k})\lambda_{1,2}({\bf k},\omega)\nonumber\\
-\frac{\left(\sum_{\bf k}\gamma_{1,2}({\bf
    k})\lambda_{1,2}({\bf k},\omega)\right)^2}{N\sum_{\bf
    k}\lambda_{1,2}({\bf k},\omega)},
\end{eqnarray}
and
\begin{eqnarray}
&&\Delta\chi(\omega)=\frac{\sum_{\bf k}\lambda_1({\bf k},\omega)\sum_{\bf
    k}\lambda_{2}({\bf k},\omega)}{N\sum_{\bf k}(\lambda_1({\bf k},\omega)+
\lambda_2({\bf k},\omega))}\nonumber\\
&&\times \left[\frac{\sum_{\bf k}\gamma_1({\bf k})\lambda_1({\bf
          k},\omega)}{\sum_{\bf k}\lambda_1({\bf k},\omega)}-
\frac{\sum_{\bf k}\gamma_2({\bf k})\lambda_2({\bf
          k},\omega)}{\sum_{\bf k}\lambda_2({\bf k},\omega)}\right]^2.
\end{eqnarray}
Thus for $A_{1g}$, the screened Raman response can be considered
as a sum of the screened response for each band, plus a mixing
term $\Delta\chi$. Here one can see that if the energy gaps and
Raman vertices are the same for each band the mixing term
vanishes, while for all other cases it is finite, reflecting the
contribution from charge backflow. In this case, the  light
scattering  induces interband charge transfer in order to recover
particle number conservation when the charge fluctuations differ
on the two bands.

As illustrated in Fig. \ref{Fig:5}, the screened $A_{1g}$ Raman spectrum
consists of contributions from each Fermi surface with a gap scale differing
by a factor of 2 and an interference contribution coming from the charge
backflow. All singularities associated with the values of the gaps at the
stationary points that would appear in the unscreened $A_{1g}$
channel are removed and replaced by cusp-like behavior as in the single band
case. However the structure of $\chi_{sc}^{\prime\prime}(\omega)$ is changed
by the interference.

Interference terms also occur if the gaps are identical on each
sheet but the Raman vertices differ. A more detailed examination
of the role of charge backflow is presented in Ref.
\onlinecite{DevVirZaw}.

\section{The Four Fermi surface Raman Spectra}
 \label{realistic}

\begin{figure}
\includegraphics[width=\columnwidth,angle=0]{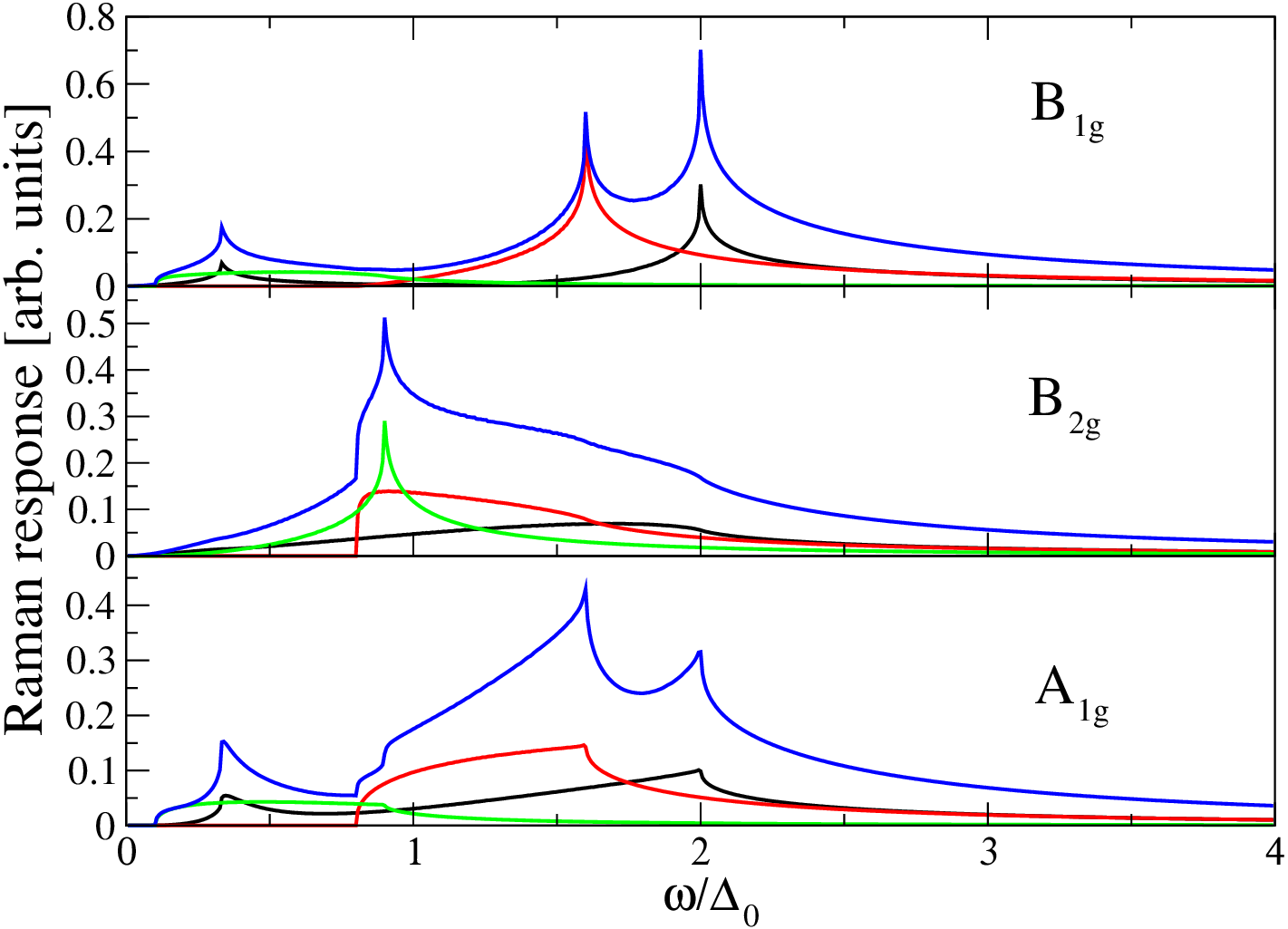}
\caption{ The $B_{1g}, B_{2g}$ and screened $A_{1g}$ Raman spectra for a
  four-sheeted Fermi surface, with $r_{\beta}=1.4$. The black line corresponds to each contribution
  from the $\beta_{1,2}$ bands, the red line to the $\alpha_1$ band, the green
  line $\alpha_2$, and the blue line is the total response. Note that for the
  $A_{1g}$ case (bottom panel) there is an interference term in addition
  to the sum of the contributions from each band. The energy gaps used are defined in the text.
 }
\label{Fig:7}
\end{figure}

\begin{figure}
\includegraphics[width=\columnwidth,angle=0]{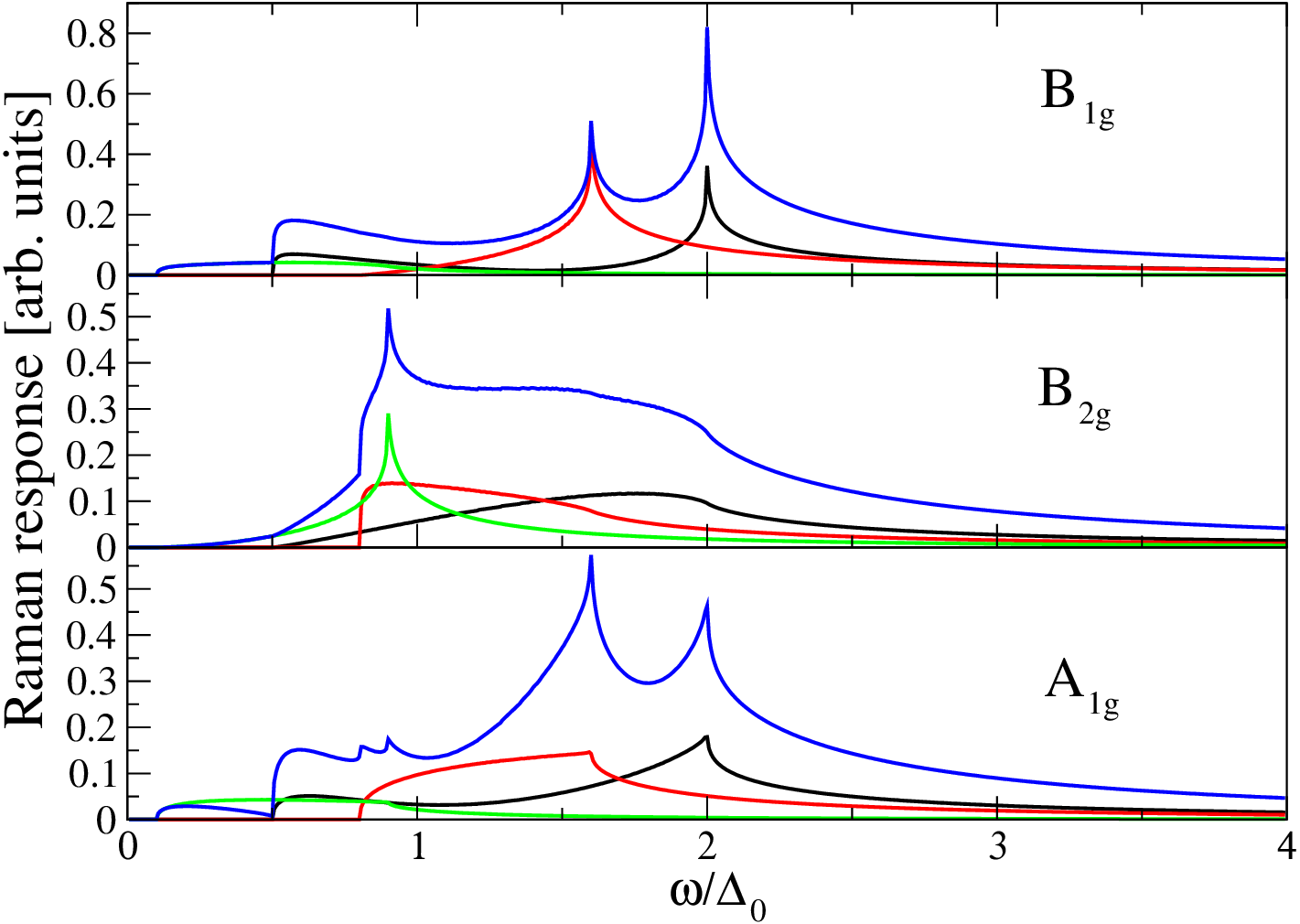}
\caption{ Same as Fig. \ref{Fig:7} but for $r_{\beta}=0.6$. }
\label{Fig:8}
\end{figure}

We would now like to extend this discussion of the features in the
Raman intensity to the 4-Fermi surface model of the Fe-pnictides.
Here we first consider the two sets of
$\Delta_{\nu}(\theta)$ gap variations for the four Fermi surface
sheets $\nu=\alpha_1, \alpha_2, \beta_1$ and $\beta_2$ shown in
Fig. \ref{Fig:4}. Specifically, the energy gaps are taken to be
(Fig. \ref{Fig:2})
\begin{eqnarray}
&&\Delta_{\beta_1,\beta_2}=\Delta_0\frac{1\mp
    r_{\beta}\cos(2\theta)}{1+r_{\beta}},\nonumber\\
&&\Delta_{\alpha_1}=-0.8\Delta_0\frac{1 +
    \cos(4\theta)}{2},\nonumber\\
&&\Delta_{\alpha_2}=-0.4\Delta_0\frac{1-\cos(4\theta)}{2}
,\label{gaps_5band}
\end{eqnarray}
As previously discussed, these gap choices are motivated by the anisotropic gaps found in
RPA spin fluctuation and functional renormalization group
calculations.  The amplitudes of the $\alpha_1$ and $\alpha_2$
gaps have been chosen to avoid an accidental overlap of
singularities between the $\alpha$ and $\beta$ gap extrema.   The
first of these, shown by the solid curves in Fig. \ref{Fig:2},
corresponds to an $A_{1g}$ gap with nodes on the $\beta$-Fermi
surfaces ($r_{\beta}=1.4$), and the second one is nodeless
corresponding to the dashed curves ($r_{\beta}=0.6$). The $B_{1g},
B_{2g}$ and $A_{1g}$ spectra for these two cases are shown in
Figs. \ref{Fig:7} and \ref{Fig:8}, respectively. The contribution
from the individual Fermi surfaces are also indicated. 

For the
$B_{1g}$ and $B_{2g}$ spectra the structure seen in the total
response are just the sum of the spectra from the individual Fermi
surface sheets with the appropriate gaps shown in Fig.
\ref{Fig:2}. For example, in Fig. \ref{Fig:7}, the $B_{1g}$
response for the gap with nodes consists of a sum over different
contributions coming from the two hole Fermi surfaces $\alpha_1$
and $\alpha_2$ and the sum of the spectra from the $\beta_1$ and
$\beta_2$ electron Fermi surface sheets which are identical since
Raman only probes $\mid\Delta\mid$ and is not sensitive to the
phase in the absence of impurities. Just as the for the previous
discussion of the single Fermi surface case, one can easily
identify the characteristic features coming from each Fermi
surface shown in Fig. \ref{Fig:7}. The $B_{1g}$ Raman response for
a gap with nodes on the $\beta$ sheets (red curve) exhibits log
singularities at $\omega/\Delta_0$ equal to
$2\Delta_{\beta,max}/\Delta_0\simeq 1.5$ and
$2\mid\Delta_{\beta}\mid_{min}/\Delta_0\simeq 0.25$. The
$\alpha_1$ sheet (green) contributes an additional log singularity
at $2\mid\Delta_{\alpha_1}\mid_{max}/\Delta_0\simeq 1.6$. Since
the $B_{1g}$ vertex has a node at the minimum value of the gap on
the $\alpha_1$ sheet the discontinuity at $\omega/\Delta_0=0.8$ is
eliminated leaving only a linear onset. The contribution of the
$\alpha_2$-sheet has a similar linear onset at $\omega=0$ and a
change in the slope at
$2\mid\Delta_{\alpha_2}\mid_{max}/\Delta_0\simeq 0.8$ due to the
fact that the $B_{1g}$ Raman vertex has a node at the minimum and
maximum values of the gap on the $\alpha_2$ Fermi surface. 

\begin{figure}
\includegraphics[width=\columnwidth,angle=0]{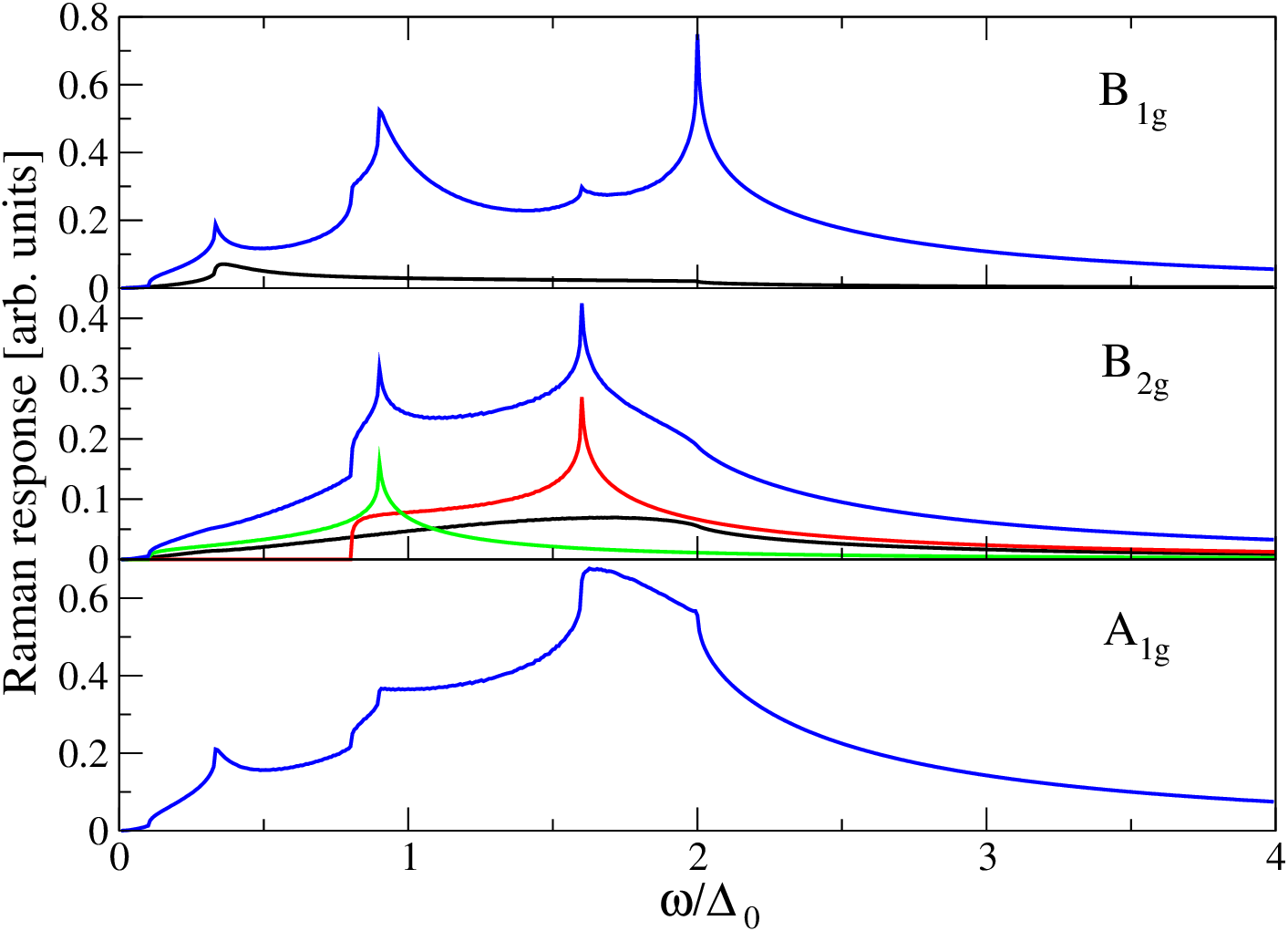}
\caption{ Same as Fig. \ref{Fig:7} but for different Raman vertices, given by
  Eq. (\ref{new-vertices}). }
\label{Fig:9}
\end{figure}

A similar discussion can be given for the $B_{2g}$ case shown in the
center panel of Fig. \ref{Fig:7}, where the total response is the
sum of the contribution from the individual Fermi surface sheets
with the $B_{2g}$ Raman vertex. As discussed in the previous
section, the screened $A_{1g}$ response shown in the lower panel
of Fig. \ref{Fig:7} contains the screened contributions from each
band, plus an interference term with interference contributions
coming from each pair of bands. All singularities are removed by
the backflow, leaving the cusp-like behavior which can be tied to
each of the gap extrema, as done for the other channels.

As mentioned in Sec. \ref{theory}, different forms for the Raman vertices are
in principle allowable other than the ones given in Eq. (\ref{vertices}), giving
the Raman spectra shown in Figs. \ref{Fig:7} and \ref{Fig:8}. Here we consider
how different forms for the Raman vertices affect the general structure of the
channel dependent Raman spectra. As an example, we consider the following form
for the Raman vertices:
\begin{eqnarray}
\label{new-vertices}
\gamma_{\alpha_1,\alpha_2}(\theta)^{A_{1g}} &=& 1\nonumber\\
\gamma_{\beta_1,\beta_2}(\theta)^{A_{1g}} &=& -1,\nonumber\\
\gamma_{\alpha_1,\alpha_2}(\theta)^{B_{1g}}  &=& \cos(2\theta)\nonumber\\
\gamma_{\beta_1,\beta_2}(\theta)^{B_{1g}}  &=& \pm 1,\nonumber\\
\gamma_{\alpha_1,\alpha_2}(\theta)^{B_{2g}}&=& \sin(2\theta)\nonumber\\
\gamma_{\beta_1,\beta_2}(\theta)^{B_{2g}}&=& \sin(\theta).
\end{eqnarray}
These vertices are all allowed by symmetry and have less anisotropy than the
ones considered in Eq. \ref{vertices} and thus highlight different gap
structures. Moreover, due to the angle-independent form for the $B_{1g}$
vertices on the $\beta$ sheets, screening must be included and the mixing terms are non-zero to give finite
contributions to the spectra. 

The resulting channel-dependent spectra are
shown in Fig. \ref{Fig:9} for the Fermi sheet gaps given in
Eq. \ref{gaps_5band} with $r_{\beta}=1.4$. For the $B_{1g}$ case, Raman
scattering from the individual $\beta$ sheets now is canceled by backflow due
to the angular independent Raman vertices on those sheets, and two identical
contributions arise from the $\alpha$ sheets, as in Fig. \ref{Fig:7}. However,
an additional mixing term arising from scattering interferences involving
each separate $\beta$ sheet, gives a large contribution to the spectra, with
peaks occuring at the energy scales given by the energy gap extrema on each
sheet. For the $B_{2g}$ spectra, the sinusoidal variation $\sin(\theta)$ of the vertices on
the $\beta$ sheets now allow for the highlighting of the gap maximum, giving a
peak frequency at twice the gap maximum for the $\beta$ sheets, in contrast to
the spectra shown in Fig. \ref{Fig:7}. A dramatic change of the spectra is
observed for the $A_{1g}$ channel. Due to backflow, the Raman response from
each separate band vanishes identically for the constant vertices in
Eq. (\ref{new-vertices}), but due to the change in sign, the entire Raman
spectra arises solely from the mixing terms. The positions of the cusp-like
feature at $\omega=2\Delta_0$ are the same as that shown in Fig. \ref{Fig:7} for a different form
for the Raman vertices, however the overall spectra lineshape is qualitatively
different. The overall structure of the lineshapes thus indicate that a proper
account of the anisotropy of the Raman vertices may be needed in order to
obtain a qualitative comparison with the experimental observed Raman
spectra. This is a topic for further study.

\begin{figure}
\includegraphics[width=\columnwidth,angle=0]{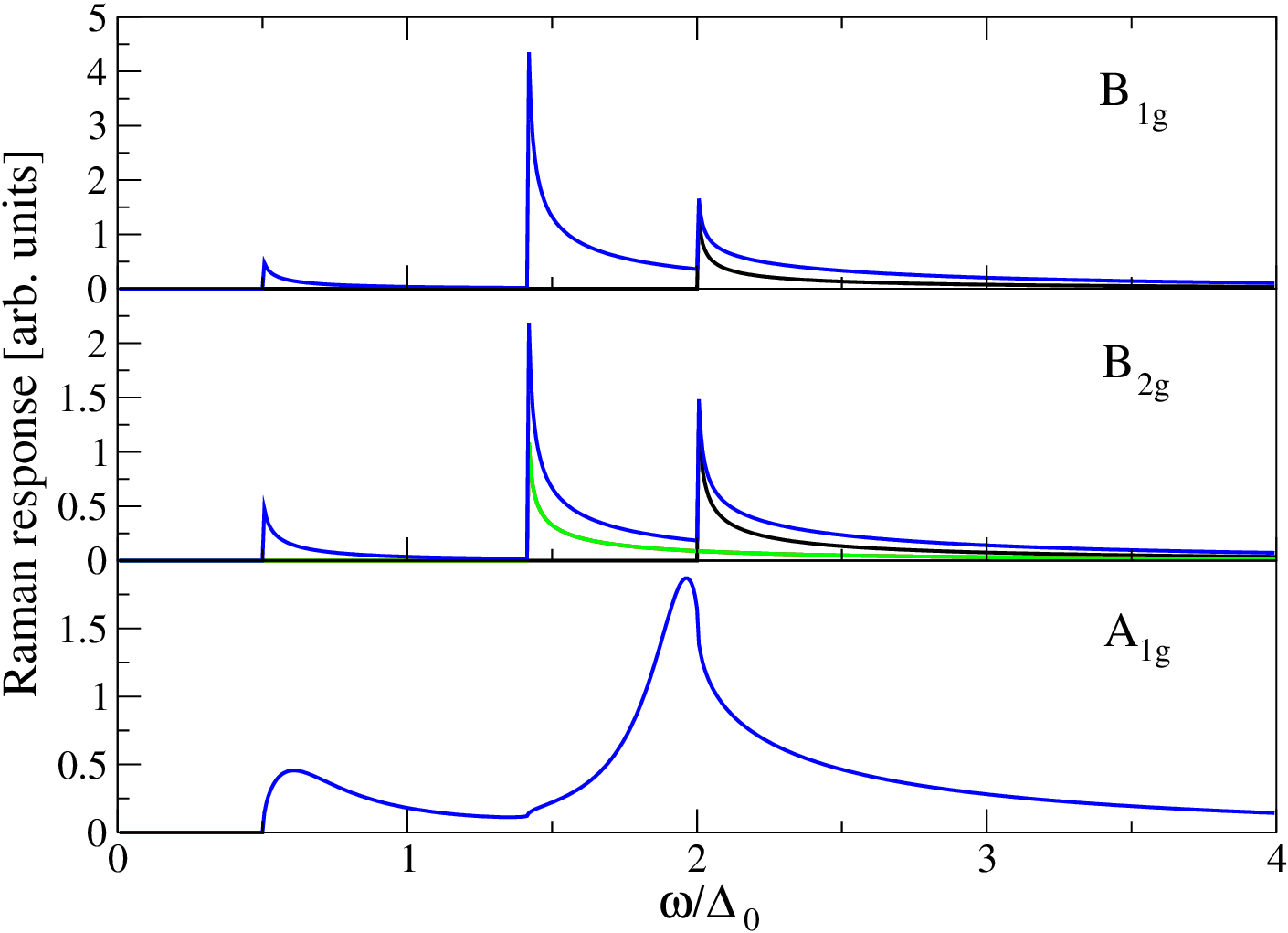}
\caption{ Polarization dependence Raman response for the $s\pm$ state, having
  energy gaps given by Eq. (\ref{S+-}) and Raman vertices given by
  Eq. (\ref{new-vertices}). Color scales, denoting contributions from each band
  and the total response, are the same as in Figs. \ref{Fig:7}-\ref{Fig:9}.}
\label{Fig:10}
\end{figure}

Finally we consider a simple sign-changing $s\pm$ state on the four sheets:
\begin{eqnarray}
\label{S+-}
\Delta_{\beta_1,\beta_2}&=&-\Delta_0/\sqrt{2},\nonumber\\
\Delta_{\alpha_1}=\Delta_0,~~&&\Delta_{\alpha_2}=\Delta_0/4.
\end{eqnarray}
Using the vertices defined by Eq. \ref{new-vertices}, the resulting Raman
spectra is shown in Fig. \ref{Fig:10}. Here, the square root divergence at
twice the gap value is ubiquitous, displaying both in $B_{1g}$ and in $B_{2g}$
The only polarization difference is that the contributions
from the $\beta$ bands for the $B_{1g}$ channel are screened, in contrast to the $B_{1g}$
channel. Apart from this difference, the spectra are qualitatively similar,
due to the mixing terms in $B_{1g}$ which restore the singularity at $\omega=1.5\Delta_0$. 
For the $A_{1g}$ channel, the divergences are screened and the
response maintains thresholds and peaks at twice the gap energy for each
band. 

\section{Conclusions}

\label{conclusions}

Here we have studied the Raman scattering response for some simple
models of the Fe-pnictide superconductors. Specifically, we
considered two different anisotropic $A_{1g}$ gaps, one with nodes
and one without nodes, on four circular Fermi surfaces. Besides
the well-known low frequency differences in the spectra for nodal
and non-nodal gaps, we found a rich set of high frequency
structures arising from stationary points of
$\Delta_{\nu}(\theta)$ on the various Fermi surfaces. 
Measurements of different polarizations may allow
one to associate particular gap structures with individual Fermi
surface sheets. If the gap has a large anisotropy as suggested by
some calculations, there will be a rich Raman spectrum for
different symmetry channels. However, if one has a relatively
isotropic gap, such as the proposed sign-switched $s-$wave, the
spectrum should be simpler and sharper since for an isotropic gap
the response has a square root singularity at $2\Delta_0$ rather
than the weaker log singularity found for an anisotropic gap.

We have also discussed some of the unusual aspects of
the Raman spectra to be anticipated in the Fe-pnictides due to
their multisheeted Fermi surface electronic structure.  In particular, Coulomb
backflow mixing may affect the $B_{1g}$ spectrum as well as $A_{1g}$. There are
other unusual aspects, such as excitonic modes associated with
short range interactions\cite{ref:Chubukov_raman} or competing
order parameter channels, which we have not explored here, but may
make the spectra in these materials even richer.  Work along these
lines is in progress.

 \acknowledgements
 We thank L.Kemper, T.Maier, S.Graser for useful discussions.
 Research was
 partially supported by DOE DE-FG02-05ER46236 (PJH) and
DOE DE-AC02-76SF00515 (TPD). DJS thanks the Stanford Institute of Theoretical
Physics and the Stanford Institute for Materials and Energy Sciences for their hospitality.


\end{document}